# An Efficient and Adaptive Framework for Achieving Underwater High-performance Maintenance Networks

Yu Gou[1,2][0000-0002-6449-2927] and Tong Zhang[1,2][0000-0002-8056-3553] and Jun Liu[2*][0000-0001-6422-2911] and Zhongyang Qi[1][0000-0003-0430-8232] and Dezhi Zheng[3][0000-0003-3998-5989]

[1] Beihang Ningbo Innovation Research Institute, Beihang University, Ningbo, China, 315800

[2] School of Electronic and Information Engineering, Beihang University, Beijing, China, 100191

[3] MIIT Key Laboratory of Complex-Field Intelligent Sensing, Beijing Institute of Technology, Beijing 100081, China
liujun2019@buaa.edu.cn

**Abstract.** With the development of space-air-ground-aqua integrated networks (SAGAIN), high-speed and reliable network services are accessible at any time and any location. However, the long propagation delay and limited network capacity of underwater communication networks (UCN) negatively impact the service quality of SAGAIN. To address this issue, this paper presents U-HPNF, a hierarchical framework designed to achieve a high-performance network with self-management, self-configuration, and self-optimization capabilities. U-HPNF leverages the sensing and decision-making capabilities of deep reinforcement learning (DRL) to manage limited resources in UCNs, including communication bandwidth, computational resources, and energy supplies. Additionally, we incorporate federated learning (FL) to iteratively optimize the decision-making model, thereby reducing communication overhead and protecting the privacy of node observation information. By deploying digital twins (DT) at both the intelligent sink layer and aggregation layer, U-HPNF can mimic numerous network scenarios and adapt to varying network QoS requirements. Through a three-tier network design with two-levels DT, U-HPNF provides an AI-native high-performance underwater network. Numerical results demonstrate that the proposed U-HPNF framework can effectively optimize network performance across various situations and adapt to changing QoS requirements.

**Keywords:** Space-air-ground-aqua integrated networks (SAGAIN), Underwater Communication Networks (UCNs), Deep Reinforcement Learning (DRL), Federated Learning (FL), Digital Twins (DT).

## 1 Introduction

In the Beyond 5G (B5G)/sixth-generation (6G) era, networks must consistently provide high-speed and reliable services at all times and in all locations. Space-air-ground-aqua integrated networks (SAGAIN) are considered one of the major enabling technologies for 6G networks, as they combine satellite communication networks, air-based networks, ground-based networks, and marine communication networks to offer wide area coverage and large network capacity [1]. However, before SAGAIN can effectively service numerous applications, the performance of maritime communication networks, particularly underwater communication networks (UCNs), must be optimized.

Recent years have witnessed the rapid growth of Internet of Underwater Things (IoUT) and machine learning (ML) [2]. Researchers are now aiming to combine ML technologies with IoUT to optimize network performance through the decision-making capabilities of intelligent algorithms, and to utilize network data to iteratively enhance these intelligent algorithms. This approach aspires to create AI-native UCNs [3][4]. While integrating intelligence into UCNs has the potential to improve network effectiveness, it also introduces several new challenges.

First, as network application requirements and scenarios become more sophisticated, the models used for underwater tasks also become increasingly complex. Training such models requires a substantial amount of data, and the transmission of this data raises communication overhead and decreases network efficiency. Second, information leakage during communication can pose a significant threat to network security and privacy, particularly in defense-related applications. Third, intelligent models are usually designed for specific circumstances or tasks and have limited generalizability. When network conditions or high-level objectives change, these deployed models will eventually lose their utility.

In addition, underwater acoustic channels are regarded as one of the most challenging communication channels. They combine the long propagation delay characteristics of satellite communications with the unstable link quality of terrestrial mobile radio networks [5]. Real acoustic communication systems may not provide sufficient training data for intelligent algorithms, making it necessary to build acoustic communication simulators. However, the quality and propagation speed of acoustic communication are affected by time-varying and non-reproducible variables such as temperature, turbulence, and ambient noise. Consequently, there is a significant sim-to-real gap between existing simulators and real environment, which leads to poor performance of intelligent algorithms.

To construct high-performance maintained underwater communication networks (U-HPNets), this paper proposes an efficient and adaptive framework, termed U-HPNF. U-HPNF is developed with a hierarchical architecture that integrates deep reinforcement learning, federated learning, and digital twin approaches to address prevalent UCNs concerns of long propagation delays, limited energy, and low channel utilization. Notably, U-HPNF takes data privacy into account, a factor rarely explored in underwater network research. The decision-making model parameters are federatively aggregated and updated through the combination of deep reinforcement learning and federated learning. Additionally, U-HPNF leverages the computational capability of nodes at all network levels to handle various network situations. The integration of deep reinforcement learning, federated learning, and digital twin techniques enhances the adaptability of intelligent algorithms in complex environments and minimizes the risks associated with directly applying intelligent algorithms to real systems. U-HPNF can adapt to the dynamic acoustic communication environment and diverse QoS requirements through iterative optimization, achieving a high-performance maintained network with self-management, self-configuration, and self-optimization capabilities.

## 2 Framework of Underwater High Performance Maintenance Networks

### 2.1 Multi-layer architecture for achieving U-HPNets

This study presents the hierarchical U-HPNF architecture designed to achieve an underwater high-performance maintained network with self-management, self-configuration, and self-optimization capabilities, as illustrated in Fig. 1.

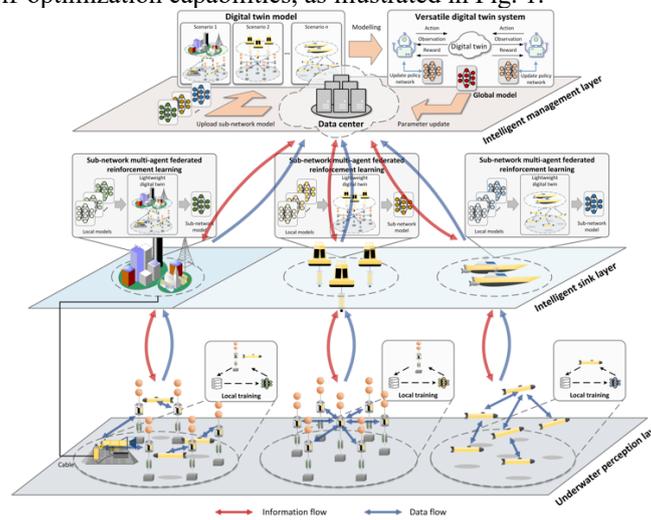

**Fig. 1.** Working diagram of U-HPNF.

U-HPNF consists of three layers: the underwater perception layer at the node level, the intelligent sink layer at the subnet level, and the intelligent aggregation layer at the network level. Perception layer comprises underwater end devices with limited battery supply and computational power, submarines, autonomous underwater vehicles. Using a local DRL-based decision-making model, the underwater node optimizes the transmission parameters to enable efficient and reliable underwater communications. The intelligent sink layer comprises buoys, surface gliders, unmanned surface vehicles (USV), and ship-based mobile platforms with long self-sufficiency and powerful computational capabilities, acting as a cross-domain gateway for sub-network level resource scheduling, data fusion, and dynamic network topology reconfiguration. The functions of the intelligent management layer are implemented in a data center, which has sufficient power supplies and computational capabilities to mimic various potential network scenarios, allowing network operators to perform online network optimization and what-if analysis in response to changes in network QoS requirements and scenarios.

Throughout the network lifecycle, the intelligent management layer provides network-level coordination, utilizing digital twin technology to analyze, optimize, and extrapolate the network to attain a high level of network autonomy.

### 2.2 Underwater Perception Layer

Each U-HPNF node consists of a perception module (P-Module), a decision-making module (DM-Module), a network module (Net-Module), and a replay buffer. The P-Module perceives the node and channel states through built-in energy, channel, request, and traffic models. The DM-Module takes the information observed by the P-Module as input and, through a DRL-based policy network, outputs a set of configurations for the Net-Module. The reply buffer stores the transitions generated by the continuous interactions between the node and the environment. The cached samples are used to update the DM-Module when the node's processing capacity and energy supply are sufficient for local training. All nodes that utilize the DRL-based decision-making module to adjust transmission parameters are referred to as intelligent nodes or agents, and all agents have the same objectives as the high-level network QoS requirements.

### 2.3 Intelligent Sink Layer

In addition to functioning as gateways and network performance optimizers, sink nodes are typically mobile and employed for reconnaissance, early warning, and maritime environmental assessment missions. Although underwater devices are also capable of updating model parameters in response to environmental changes, their ability to deal with network topology changes and node failures is constrained by computational power, energy, and partial observability. Consequently, intelligent sub-network level management requires surface sink layers. The lightweight network digital twin of U-HPNF is deployed in the intelligent sink layer. To safeguard the security and privacy of the underwater network and prevent unnecessary data transmission, only model parameters are transferred in U-HPNF without providing local node observations and the model structure. Sink nodes can update node model parameters and preserve subnet performance when node failures or subnet topology changes are observed by mimicking several conceivable situations in lightweight network digital twins. Model updates in the intelligent sink layer can be considered as an extension of the node-level independent q-learning (IQL) paradigm to the centralized training with decentralized execution (CTDE) paradigm.

### 2.4 Intelligent Aggregation Layer

Network-level model modifications are necessary in the data center when network requirements change. In order to shorten the time required to revise the configuration of the intelligence algorithms, network operators generally train optimization models for potential scenarios in advance, which requires additional storage capacity. In addition to distributing the modified subnet-level model parameters to the related underwater nodes, the sink node also sends the model parameters to the data center. With the obtained parameters, the data center updates the model of the virtual nodes in the associated subnetwork. The data center receives all joint behavior rules transmitted by sink nodes and combines all subnetwork joint policies into a network joint behavior policy.

## 3 Performance Evaluation

This section presents the numerical results for U-HPNF. We evaluated U-HPNF with three baseline methods in two typical cases. In the first case, the network consists of five underwater nodes that may encounter failures due to hardware or software malfunctions. All nodes collaborate to maximize the number of concurrent communications. In the second scenario, the network size changes, and the high-level scheduling objectives shift from maximize network capacity to optimize communication fairness. Ablation experiments compare the performance of U-HPNF and its variants and validate that U-HPNF adapts under changing network settings. Section 3.2 demonstrates that U-HPNF is able to maintain network performance when node encounters failure, and Section 3.3 confirms that U-HPNF can cope with various network high-level objectives and topologies.

### 3.1 Evaluation Setup

Consider an underwater network with $N$ underwater nodes, where transmitters are evenly distributed at the bottom of the cylinder region, and receivers are located at the top. The cylinder region has a radius of 4 km and a height of 1 km. For the application requirements, the duration of each transmission slot is set to ten seconds in both cases. We implemented the DRL-based decision-making model with two fully connected layers and a GRU. The first fully connected layer comprises of 64 hidden units, followed by a GRU with 64 hidden units and a ReLU to process the observations. The second fully connected layer comprises seven hidden units that generate Q-values for each available action. For training the DRL-based decision-making model, we set the replay buffer size $\mathcal{B}_r$ to 10,000 and the mini-batch size $b$ to 32. The model was trained over 300,000 episodes, with the target network updated every 200 episodes. We employed an $\varepsilon$-greedy behavior policy to balance exploration and exploitation, with $\varepsilon$ decreasing linearly from 1 to 0.05 for the first 150,000 episodes and remaining constant at $\varepsilon$=0.05 for the last 150,000 episodes. The discount factor was set to 0.99. The available transmit power set is [0, 2, 4, 8, 16, 64] W, where 0 W indicates that the node is not scheduled in current transmission slot.

### 3.2 Subnet-level performance optimization

In a subnet with five nodes, all nodes collaborate to maximize the number of concurrent communications. Each underwater node is identical, equipped with the same battery capacity and acoustic modems. The nodes independently select the transmission power based on their perceptions of the environment and their own states, and they transmit environmental parameters to the intended receivers. It is assumed that each underwater node may fail to load the intelligent algorithm with a specific probability $\epsilon$ and will then transmit at random power level, regardless of the quality of other communications or the overall network performance. Node failures follow a Bernoulli distribution.

Figure 2 compares U-HPNF with four typical network management algorithms and four variants as $\epsilon$ increases from 0 to 0.2 in the aforementioned network topologies: 1) *Greedy* [6]: All nodes transmit simultaneously and operate at the maximum available power for higher throughput.; 2) *TDMA*: Nodes transmit at the maximum available power during their allocated slot; 3) *Random* [7]: Nodes transmit at arbitrary power levels and may refrain from transmitting during some time slots (by selecting a power level of 0 W); 4) *IQL* [8]: A Q-learning-based joint link scheduling and power allocation algorithm designed to alleviate interference in underwater communications, thereby improving network capacity. *w/o failure* refers to the model trained in perfect UCNs, and *w/i 0.01*, *w/i 0.1,* and *w/i 0.2* refer to the models trained in UCNs with fixed failure rates of 0.01, 0.1, and 0.2, respectively. *w/i responsive* represents the model periodically updated with the sink layer in U-HPNF. The *y*-axis denotes the average number of concurrent communications at each time slot across 60 executions. Specifically, Greedy, TDMA, and Random are static strategies that do not respond to changes in the communication environment or QoS requirements. IQL seeks the optimal transmission parameters for individual nodes, whereas U-HPNF ant its variants optimize network performance through coordination.

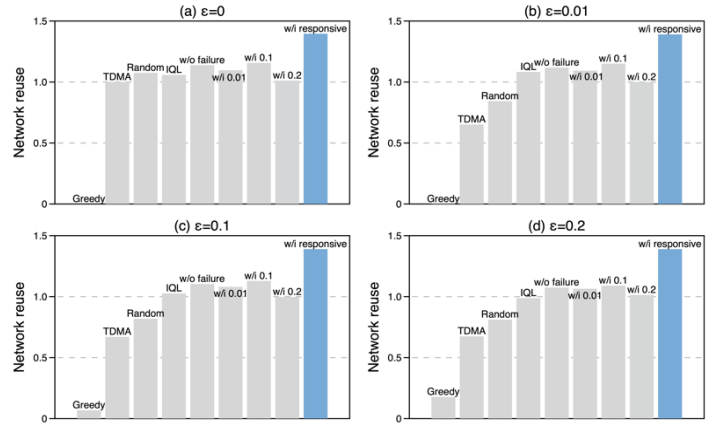

**Fig. 2.** Comparisons of U-HPNF and other intelligent network management algorithms when (a) $\epsilon$=0, (b) $\epsilon$=0.01, (c) $\epsilon$=0.1, and (d) $\epsilon$=0.2.

Greedy fails to optimize network performance when all nodes transmit simultaneously at the maximum power level. As $\epsilon$ gradually increases, more nodes degrade into failure nodes, leading to higher network reuse with the Greedy algorithm. This result indicates that Greedy may not be suitable for dense networks. TDMA allocates equal transmission opportunities to all transmitters and achieve a network reuse of 1 when $\epsilon = 0$. However, as some nodes gradually lose functionality, network reuse declines by 35%, demonstrating that TDMA lacks robustness in real-world deployments. Random shows better robustness than TDMA. The comparison among IQL, U-HPNF, and its four variants reveals several insights. First, IQL optimizes network performance by maximizing the performance of individual nodes, whereas U-HPNF coordinates the transmission behaviors of all nodes. The numerical results indicate that coordination provides significant benefits in multi-user systems. Second, models trained in simulators with nodes failures (i.e., *w/i 0.01*, *w/i 0.1,* and *w/i 0.2*) do not perform better than those trained without failures (i.e., *w/o failure*), highlighting that generalizability is crucial for optimizing network performance. These evaluations also validate the efficacy of deep reinforcement learning in optimizing network performance across dynamic, complicated scenarios.

### 3.3 Network-level performance optimization

Consider two scenarios with different high-level scheduling objectives: the first aims to maximize network capacity, and the second aims to optimize communication fairness. When the network scenario changes, the data center transmits appropriate decision-making model to the underwater nodes. Model I is used to maximize network capacity, whereas Model II is used to optimize network fairness. The *y*-axis in Fig. 3(a) and (b) show the achieved network capacity (in kb) and communication fairness during 60 time slots, respectively. Network capacity is calculated using the Shannon formula as described in [3], and communication fairness is as defined according to [9]. The *x*-axis in both subfigures represent the number of transmitters.

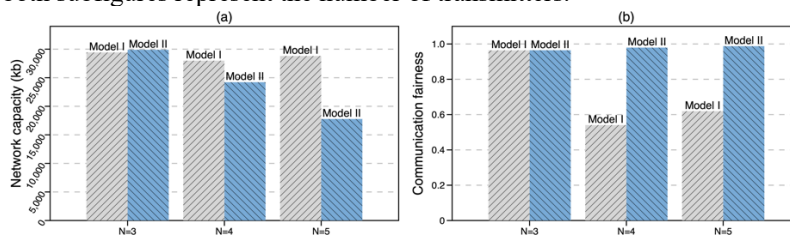

**Fig.3.** Effects of U-HPNF in handling various high-level network objectives: (a) maximize network capacity and (b) optimize communication fairness.

Figure 3 reveals that Model I and II have comparable network capacity and communication fairness at $N = 3$ because link scheduling and power control allow for more concurrent communication in the network when fewer network nodes are deployed. However, as the number of nodes in the network increases to four, the communication fairness of Model I decreases sharply, although it maintains its superiority in terms of network capacity. Model II presents a similar trend. While Model II is able to preserve communication fairness as the number of nodes increases, its network capacity gradually declines (only 17,779.38 kb at $N = 5$, which is 40% less than that of Model I).

There are two key takeaways from these results. First, the intelligent sink layer is capable of adapting to changes in topology. Model I preserves network capacity, whereas Model II ensures communication fairness as the number of nodes increases. Second, when the network's high-level scheduling objectives change, the corresponding optimization model must be retrained to account for the new objectives. The intelligent sink layer cannot maintain optimal network performance across all possible circumstances.

## 4 Conclusion

This paper presented U-HPNF that aims to establish high-performance maintained UCNs with self-management, self-configuration, self-optimization, and self-healing capabilities. U-HPNF integrates a variety of cutting-edge machine learning approaches into underwater communication networks to optimize network performance in dynamically changing communication environments, protect network security and information privacy, and enable online data-driven network performance optimization and model training. Simulation results confirmed the efficacy of integrating deep reinforcement learning, federated learning, and digital twin techniques for maintaining high performance in networks and scenarios that are dynamically changing. Additionally, these results underscore the necessity of the U-HPNF layered structure. We hope that U-HPNF will enable high-speed and reliable SAGAIN in the B5G/6G era.

## Acknowledgement

This work was supported in part by the National Key Research and Development Program (No. 2021YFC2803000), the National Key Basic Research Program (No. 2020YFB050001), the National Natural Science Foundation of China (No. 62101211 and No. 61971206), and the Joint Funds of the National Natural Science Foundation of China (No. U22A2009).

## References

1. Hongzhi Guo, Jingyi Li, Jiajia Liu, Na Tian, and Nei Kato. A survey on space-air-ground-sea integrated network security in 6g. IEEE Communications Surveys & Tutorials, 24(1):53–87, 2021.
2. Amal Feriani and Ekram Hossain. Single and multi-agent deep reinforcement learning for ai-enabled wireless networks: A tutorial. IEEE Communications Surveys & Tutorials, 23(2):1226–1252, 2021.
3. Tong Zhang, Yu Gou, Jun Liu, Tingting Yang, and Jun-Hong Cui. Udarmf: An underwater distributed and adaptive resource management framework. IEEE Internet of Things Journal, 9(10):7196–7210, 2021.
4. Yu Gou, Tong Zhang, Tingting Yang, Jun Liu, Shanshan Song, and Jun-Hong Cui. A deep marl- based power management strategy for improving the fair reuse of uwsns. IEEE Internet of Things Journal, 2022.
5. Milica Stojanovic and James Preisig. Underwater acoustic communication channels: Propagation models and statistical characterization. IEEE communications magazine, 47(1):84–89, 2009.
6. Yu, W., Chen, Y., Tang, Y., & Xu, X. (2018, September). Power allocation for underwater source nodes in uwa cooperative networks. In 2018 IEEE international conference on signal processing, communications and computing (ICSPCC) (pp. 1-6). IEEE.
7. Lee, H. W., Modiano, E., & Le, L. B. (2011). Distributed throughput maximization in wireless networks via random power allocation. IEEE transactions on mobile computing, 11(4), 577-590.
8. Wang, H., Li, Y., & Qian, J. (2019). Self-adaptive resource allocation in underwater acoustic interference channel: A reinforcement learning approach. IEEE Internet of Things Journal, 7(4), 2816-2827.
9. Rajendra K Jain, Dah-Ming W Chiu, William R Hawe, et al. A quantitative measure of fairness and discrimination. Eastern Research Laboratory, Digital Equipment Corporation, Hudson, MA, 21, 1984.